# V0LTpwn: Attacking x86 Processor Integrity from Software

*Zijo Kenjar[1], Tommaso Frassetto[1], David Gens[2], Michael Franz[2], and Ahmad-Reza Sadeghi[1]*

[1]*Technische Universität Darmstadt, Germany*
{zijo.kenjar,tommaso.frassetto,ahmad.sadeghi}@trust.tu-darmstadt.de

[2]*University of California, Irvine*
{dgens,franz}@uci.edu

## Abstract

Fault-injection attacks have been proven in the past to be a reliable way of bypassing hardware-based security measures, such as cryptographic hashes, privilege and access permission enforcement, and trusted execution environments. However, traditional fault-injection attacks require physical presence, and hence, were often considered out of scope in many real-world adversary settings.

In this paper we show this assumption may no longer be justified. We present V0LTpwn, a novel hardware-oriented but software-controlled attack that affects the integrity of computation in virtually any execution mode on modern x86 processors.[1] To the best of our knowledge, this represents the first attack on x86 integrity from software. The key idea behind our attack is to *undervolt* a physical core to force non-recoverable hardware faults. Under a V0LTpwn attack, CPU instructions will continue to execute with erroneous results and without crashes, allowing for exploitation. In contrast to recently presented side-channel attacks that leverage vulnerable speculative execution, V0LTpwn is not limited to information disclosure, but allows adversaries to affect execution, and hence, effectively breaks the integrity goals of modern x86 platforms. In our detailed evaluation we successfully launch software-based attacks against Intel SGX enclaves from a privileged process to demonstrate that a V0LTpwn attack can successfully change the results of computations within enclave execution across multiple CPU revisions.

## 1 Introduction

Modern hardware platforms have a long history that spans multiple decades. The need to ensure backwards compatibility and the constant tweaking of existing designs has burdened widely deployed hardware architectures with legacy components that have become highly complex, and far from flawless. In the recent past, we have seen how seemingly minor implementation bugs at the hardware level can have a severe impact on security [14]. Attacks such as Meltdown [36], Spectre [33], Foreshadow [55], and RIDL [59] demonstrate that attackers can exploit these bugs from software to bypass access permissions and extract secret data.

Furthermore, we have seen that the adverse effects of hardware vulnerabilities are not limited to confidentiality, but can also compromise integrity in principle: the infamous Rowhammer bug [32] resulted in numerous exploits [6, 26, 42, 46, 48, 54, 57, 60, 62] leveraging bit flips in flawed DRAM modules, which are deployed on practically all computer systems today. While initial defenses have been proposed to mitigate Rowhammer from software [5, 8], fixing hardware bugs ultimately requires deploying new hardware.

With recent feature sizes shrinking to single-digit nanometer scale, semiconductor companies face the growing problem of the so-called *dark silicon*. At run time large parts of the chip will have to be left *powered-off*, since the billions of transistors cannot be operated within the thermal constraints and power budget the platform was originally designed for. This prevented hardware designers from leveraging Dennard scaling [17, 51]; consequently, manufacturers have moved to more intelligent, on-demand thermal and voltage control on recent platforms. This means that critical operational aspects of the processor can now and are increasingly controlled from software during run time. Unfortunately, this development comes with severe consequences for computer security.

In 2017 Tang et al. [53] showed that the intricacies of low-level and fine-grained power management on ARM-based mobile devices open up serious pitfalls, as they were able to induce faults in the processor of a Nexus 6 smartphone, allowing them to bypass the isolation boundary of TrustZone. So far, a similar scenario was deemed unlikely on x86-based systems for several reasons: (i) x86-based power management traditionally does not expose direct access to hardware regulators to software above the BIOS level, (ii) desktops and servers are typically not battery powered, and hence, feature less aggressive and more coarse-grained power management, and finally (iii) x86-based platforms deploy extensive safety measures and implement strict architectural defenses to prevent, detect, and recover from hardware faults at run time. We elaborate on the differences between our work and previous attacks in Section 8.

---

[1]We responsibly disclosed the findings of this paper. Intel assigned CVE-2019-11157 to the vulnerability and also issued an advisory.

In this paper, we present V0LTpwn, the first software-controlled fault-injection attack for x86-based systems. Our attack is able to directly affect processor execution regardless of privilege level, execution mode, or hardware isolation. As a result, V0LTpwn is also able to compromise the integrity guarantees of Intel's Software Guard Extensions (SGX). SGX is a hardware security extension which Intel promotes in cloud-based scenarios where cloud providers should be considered untrusted [29].

The key idea behind our V0LTpwn attack is to *undervolt* the physical target core that executes the victim software (i.e., reduce its available voltage). We achieve this by exploiting software-exposed but obscure power-management interfaces of modern x86 platforms. We analyze a number of CPUs of different Intel generations and we show that all of them are prone to fault-injection attacks despite deploying dedicated counter measures. In particular, all of these processors feature an elaborate set of management and safety mechanism called Machine-Check Architecture (MCA) [30], which provides detection and fallback routines for handling critical hardware events such as core, uncore, interconnect, bus, parity, and cache errors.

Processors leverage a number of model-specific registers to control and report such events across different hardware layers. These events can then be forwarded as machine-check exceptions to software handlers to store, process, and react to critical failures. However, we show that an adversary can still inject exploitable hardware faults by carefully driving processor execution into unstable voltage domains. We construct a proof-of-concept exploit in which the attacker injects such faults into a running SGX enclave entirely from software. We analyze, conduct, and evaluate this new attack through a number of tests across multiple Intel CPUs.

Contrary to recent hardware-oriented attacks such as Foreshadow [55], Spectre [33], RIDL [59] and Meltdown [36] — which are limited to extracting information through side channels — our attack enables an adversary to *manipulate* enclave execution and compromise its integrity. Through concurrent use of execution units and by leveraging power-intensive instructions we provoke resource contention which results in reliable and reproducible faults in our tests. For this, we leverage undocumented features, extending and customizing the available software tools to enable detailed probing and attacks on real-world code. Our findings show that the deployed defenses (MCA, SGX isolation) are insufficient in practice, leaving a large number of real-world system vulnerable to V0LTpwn.

To summarize, our contributions include the following:

- **Novel attack against x86 processors:** we present *V0LTpwn*, the first software-controlled fault-injection attack for the x86 platform. Through targeted *undervolting* from malicious software V0LTpwn is able to alter computational results and affect processor execution in victim software at run time. We introduce several new techniques, such as identifying fault-susceptible frequency settings, instruction patterns, and stressing the logical partner core to increase temperature and resource contention while undervolting.

- **Real-world impact and responsible disclosure:** we confirmed reproducible and exploitable faults for code running within user processes, kernel code, and SGX enclaves. Intel confirmed our findings, assigned a CVE to this issue, and is working on a mitigation.

- **Extensive evaluation and proof-of-concept implementation:** we implement and demonstrate an end-to-end exploit against recent processors that support SGX, which is designed as a completely isolated and trusted execution environment in the presence of potentially malicious software running on the platform. By undervolting the processor while the SGX enclave runs we are able to manipulate its execution at run time and demonstrate manipulation of computation through software-induced faults. Our results show that we are able to induce and exploit faults on multiple processors of different micro-architectures despite extensive defensive measures to prevent, detect, and recover from such errors.

## 2 Background

In this section we explain the background information required for the understanding of the rest of the paper. First, we describe the principles of power management on modern x86 processors. Second, we explain undocumented software interfaces for overclocking. Third, we discuss Intel's Machine Check Architecture. Finally, we briefly cover the basics of Intel SGX.

## 2.1 Dynamic Voltage and Frequency Scaling on the x86 platform

The performance and power consumption of processors depends on frequency and voltage settings. For different software workloads, modern processors incorporate technologies for *Dynamic Voltage and Frequency Scaling (DVFS)*. In this context, processor vendors often define *performance states* (P-states), which represent distinct pairs of voltage level and clock frequency.

On recent Intel processors, DVFS techniques are included in its *Enhanced Intel Speedstep Technology (EIST)*. EIST implements hardware control of P-states and considers workload, sensor measurements, power constraints as well as software hints when selecting P-states at run time. For configuration and hints, a software interface is provided using *Model-Specific-Registers (MSR)* [28], which require supervisor privileges. Hardware control of P-states can be deactivated, for instance, to allow an operating system driver to manually transition the platform to a different P-state. In Intel's Software-Developer Manual [28], a P-state is called a *ratio*, i.e., an 8-bit value determining the frequency when multiplied with a base clock of (typically) 100 Mhz. In this paper, we will refer to P-states with the hexadecimal representation of the ratio. For instance, P-state 0x20 (i.e., decimal value 32) represents a frequency of 3200 MHz.

Since the Skylake microarchitecture Intel introduced *Hardware-Controlled Performance States (HWP)*. HWP offers a more fine-grained interface, i.e., the OS can define operation ranges for high-performance and energy-saving phases. In general, P-state definitions are model-specific as the matching core voltage for a particular frequency is defined by the hardware and may also be adjusted dynamically by the voltage regulators of the processor at run time.

## 2.2 Overclocking Interfaces

Overclocking is a common operation used to maximize processor performance on x86 processors. For an enthusiast market, manufacturers release custom *unlocked* processor models. Paired with a suitable mainboard, users are able to adjust settings like clock multiplier, voltage levels and power limits via the interfaces of the BIOS/UEFI implementation.

As a recent development, Intel has exposed traditional BIOS features to the operating system to enable *real-time overclocking*. For instance, Intel's Extreme Tuning Utility (XTU) as well as ThrottleStop allow users to adjust overclocking settings like voltage levels without a reboot of the system under Microsoft Windows. Reverse engineering has revealed the use of MSR *OC Mailbox (0x150)* by these applications. Interestingly, the official documen-tation does not disclose this functionality. However, we find references in drivers [1], presentations [44] and many mainboard manuals. We assume Intel keeps this functionality undocumented, because voltage manipulation can easily damage the hardware, and hence, requires extreme caution when applied from software.

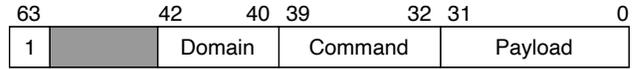

Figure 1: MSR OC Mailbox (`0x150`) is used to adjust voltage levels from software, including applications such as Intel's Extreme Tuning Utility (XTU) and ThrottleStop.

To the best of our knowledge, MSR OC Mailbox (`0x150`) has the structure depicted in Figure 1. Bit [63] is fixed and must be set to 1 in all writes to this MSR. Bits [42:40] represent a hardware domain which is addressed by the command in bits [39:32]. The lower 32 bits have a variable structure and contain the command payload. An important feature of MSR 0x150 is the ability to modify voltages. For instance, a voltage offset can be applied to the base voltage of a P-state. We found this feature to be available on all recent Intel processors. The actual voltage can be changed with 5 mV granularity. This behavior conforms to voltage regulator specifications [47], in which the voltages requests from the processor to the regulator unit are encoded in 5 mV steps. The available set of commands appears to be dependent on the microarchitecture [44]. An extended description of commands is provided in Appendix A.

## 2.3 Intel's Machine-Check Architecture

Semiconductor manufacturers achieve feature sizes within single-digit nanometer scales while continuously decreasing power-consumption per transistor to scale up performance of the chip. Unfortunately, this also causes these platforms to be increasingly sensitive to environmental conditions, such as heat and electro-magnetic radiation. This means that random hardware errors are *expected* given sufficient uptime of a running system [35]. For this reason, modern processor hardware features a set of intricate error-handling mechanisms to detect, correct, and potentially recover from such situations. One of these mechanisms is the *Machine-Check Architecture (MCA)*, which was introduced by Intel starting with the P5 architecture. MCA continuously monitors individual hardware elements, such as cores, caches, interconnects and buses, integrated controllers, etc., in real-time and logs and reports any hardware-level error conditions to a set of well-defined registers. MCA offers a programmable interface which enables system software to configure and handle trigger events based on the generated alerts.

Since serious error conditions may not allow system software to conduct any recovery (e.g., through controlled shutdown), MCA supports additional recovery options through external devices. However, since this mode of operation requires additional, non-standard setup we focus on system-level recovery using MCA in this paper. In the case of Linux and Windows the OS incorporates a driver that interfaces with the MCA registers and error handlers. Error conditions can then be logged, reported, and handled through a particular class of software interrupts, called *Machine-Check Exceptions (MCEs)*. Throughout our experiments we leveraged MCEs to aid in identifying and reverse engineering vulnerable code patterns. It is noteworthy to mention that V0LTpwn injects non-recoverable error conditions which cannot be corrected from system software, and hence, bypasses MCA.

## 2.4 Intel Software Guard Extensions

Intel's Software Guard Extensions (SGX) [29] allow developers to design hardware-protected areas, known as *enclaves*, that contain sensitive code. Access to enclaves is only allowed through specific entry points, known as *ecalls*. Unauthorized access to SGX memory, known as *Enclave Page Cache*, is disallowed by the processor. Bus snooping attacks, which consist in physically monitoring the memory bus to extract memory values, are mitigated through the use of memory encryption and memory integrity techniques. SGX offers local and remote attestation services.

SGX does not address side-channel attacks by design, leaving to the developer the burden of developing side-channel resilient code. Consequently, there have been a number of works on side-channel and micro-architectural attacks [9, 22, 23, 37, 56, 61], and side-channel defenses [4, 7, 11, 24, 45, 50, 52]. Critically, SGX does not protect against undervolting attacks either, thus allowing V0LTpwn.

To the best of our knowledge, no previous work managed to violate the integrity of computation in an SGX enclave without resorting to software vulnerabilities.

## 3 The V0LTpwn Attack

In this Section we present the main principles of our V0LTpwn attack, which injects faults in SGX enclaves by undervolting the processor.

## 3.1 Adversary Model and Assumptions

Our adversary model and assumptions are consistent with the SGX threat model. We assume:

**Root access** The attacker has control over a user process with root privileges. This also enables an adversary to query the target system, e.g., to learn the exact model number of the processor.

**DVFS** The attacker has access to software-controlled dynamic frequency scaling; all recent Intel x86 processors support it using EIST [28] (see Section 2.1). Moreover, we require the firmware to allow access to MSR 0x150, which was the case for all machines we tested.

**Target binary** The attacker has a copy of the intended victim program binary for offline testing. This is a common scenario in attacks against a well-known program or algorithm (e.g., crypto).

Unlike traditional fault-injection attacks, V0LTpwn requires no physical access to the target machine. Finally, V0LTpwn does not rely on any software vulnerabilities, and hence we do not need to make any specific assumption about the security of the code running on the platform (all code can be protected by defenses such as control-flow [3] and data-flow integrity [10], or even formally verified).

The goal of the attacker in this setting is *to tamper with the integrity* of the code executing inside an SGX enclave. While loading attacker-controlled code by corrupting SGX's setup process might be viable, we note that the impact of malicious enclaves is actually limited since enclaves are completely isolated from each other. Hence, influencing execution of benign enclaves might often be more valuable for an adversary.

## 3.2 Challenges

To implement our attack V0LTpwn, we face the following challenges:

**Symmetric Architecture** Commodity multi-core processors from Intel maintain a single voltage domain that is shared between all physical cores of the system, unlike ARM cores which can be regulated independently. As a result, undervolting the core where the victim code executes also undervolts the core running the exploit, leading to potential faults in the exploit code as well. We tackle this challenge in V0LTpwn by partitioning cores and minimizing noise throughout the system (see Section 4.1).

**Processor Diversity** Intel's x86 processors are available for different markets ranging from laptops up to high performance server systems. Although the microarchitecture is the same, these processor models are operated with different voltage levels. We address this challenge in V0LTpwn by conducting a dedicated, offline analysis phase, for which we developed a reproducible lab setup that allows us to apply attack parameters inferred from a different (but similar) physical machine to the victim machine (see Section 4.2).

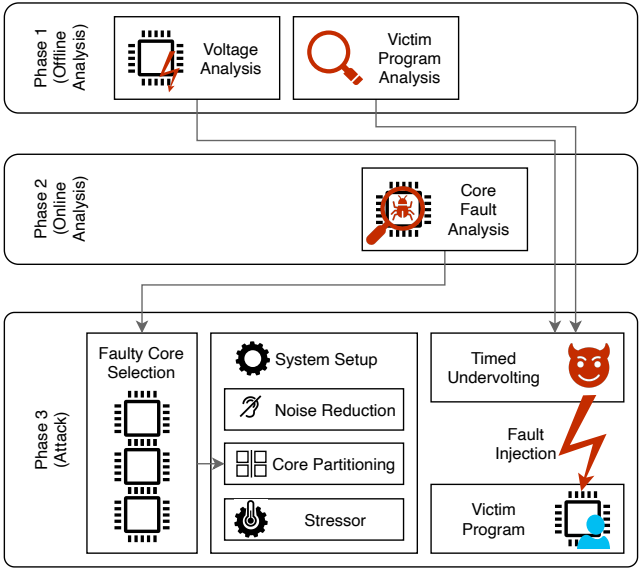

Figure 2: Overview of the V0LTpwn attack.

**Error Correction** Unlike ARM processors, Intel processors integrate the Machine Check Architecture (MCA), which is able to correct errors which occur due to undervolting [40], as explained in Section 2.3. Our attack bypasses MCA by generating non-recoverable faults (see Section 4.3).

**Undocumented Interfaces** The hardware interfaces to adjust the voltage (Section 2.2) are undocumented. To use them, we had to rely on third-party reverse-engineered partial documentation and piece it together to develop a real-world setup running on our systems, which required substantial effort on our part.

### 3.3 Attack Workflow

As mentioned before, the goal of the attacker is to exploit hardware glitches in an undervolted processor to influence the execution of an SGX enclave in a controlled way. For this, the attacker needs information about the victim's binary as well as the response to undervolting of the target processor model. Both of them can be collected offline, without interacting with the target system (*Phase 1* in Figure 2). Afterwards, the attacker needs to collect information about the physical cores in the target system, to detect which core is more prone to faults (*Phase 2* in Figure 2). With the information from Phases 1 and 2, the attacker can choose the most appropriate core in the system and mount the attack (*Phase 3* in Figure 2). We will explain these phases in the following and describe them in detail in Section 4.

**Phase 1: Offline Analysis** The attacker aims to determine a voltage level low enough to generate glitches without completely disrupting the operation of the CPU (*exploitable voltage window*). In order to determine an exploitable voltage window, the attacker progressively reduces voltage levels until faults occur, but the system does not freeze yet. During this test, the machine is likely to freeze or crash multiple times, which might be detected, if the test is performed on the target machine directly. Since the exploitable voltage window is very similar between processors of the same model, the attacker can acquire another processor of the same model and perform these initial tests on it.

Moreover, the attacker should minimize the duration of undervolting to prevent crashes on the target machine. Hence, the attacker analyzes the target binary, in order to identify parts of the code most vulnerable to faults. To this end, the attacker can scan the binary for instances of known vulnerable patterns, which we describe in Section 4.3. Next, the attacker observes the execution of the target program on the attacker's identical processor, in order to estimate at which point of the execution the binary will run the fault-prone code and for how long.

**Phase 2: Online Core Fault Analysis** In Phase 2, the attacker sets up the target system for undervolting and then probes each available core, one at a time, to determine the specific fault patterns of that core. As an example, the attacker can check how frequently the core under test experiences faults under various test conditions. This test must be done on the actual target machine, since every physical core produces different glitches while undervolted.

**Phase 3: Attack** In the previous Phases, the attacker has learned which code can be faulted and which system conditions are required to induce the fault. The attacker is now able to use this knowledge to set up the system, start the target enclave, and undervolt the processor while the enclave is running the desired code to provoke glitches in the data, thus violating the integrity of the execution.

**Target System Setup** The target platform needs to be configured in a fault-prone configuration, using the safe undervolting levels learned in Phase 1. Besides controlling the voltage, the attacker needs to limit all sources of noise, since the attack requires carefully balancing the voltage level slightly above the critical threshold to push it into fault-inducing territory at the right moment in time. Since unexpected events during this critical period can easily result in crashes or freezes, we organize processes such that the victim enclave is running alone on a core and disable various automatic management features of the hardware. This way, the victim enclave runs alone, with minimal interference, on a core of the attacker's choice, e.g., the most fault-prone.

Moreover, the attacker can further tweak the configuration of the processor to improve the performance of the attack. One option is to vary the temperature of the core, e.g., by running stressing code until the desired temperature is reached. Additionally, the attacker can run especially crafted code *(stressor)* on the logical partner of the core where the victim is executing, in order to maximize resource contention.

## 4 Implementation

This section presents our systematic approach to identify vulnerable conditions on Intel processors. First, we outline the testing procedure we developed to test for software-inducible faults on recent x86 platforms. Then we present how we identified vulnerable code patterns that yield reproducible bit flips on both *Kaby Lake* and *Coffee Lake* processors we tested in our lab.

### 4.1 Attack Setup

To ensure reproducible results and prevent interference from the run-time environment (i.e., *noise*) we first establish a setup in which disturbances from hardware and software are reduced to a minimum (or ideally, completely disabled). In the following, we explain the individual steps to achieve that.

**Controlling Voltage and Frequency** On Intel processors, the voltage and frequency are determined by the selected P-state of the cores. As the attacker, we can control them via the EIST or HWP interfaces (see Section 2.1). As a first step, we disable the operating system drivers which communicate with them. For Linux this means disabling the modules acpi_cpufreq and intel_pstate.

Second, we disable automatic hardware-based selection of P-states. In EIST, we have to set bit 0 of MSR 0x1AA to 1, which enables us to set the P-state directly using MSR 0x199. A P-state can alternatively be enforced using HWP instead of EIST (e.g., if the firmware enables it). This can then be achieved by setting the minimal, maximal, and desired P-state in MSR 0x774 to the same value. Once a P-state is set, all cores of the system are running at the same voltage level and clock frequency. Small differences are measurable because the on-die power regulation conducts small adjustments based on sensor feedback and workload [2]. Having fixed a P-state, we are now able to control the voltage levels by sending commands via MSR 0x150 (OC Mailbox).

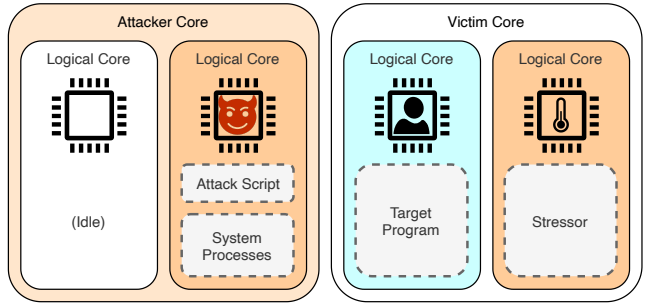

Figure 3: Core partitioning for V0LTpwn, in order to minimize noise and maximize resource contention on the target program.

**Core Partitioning and Noise Reduction** To ensure that the targeted core only runs the target application — with minimal interference — we partition all logical cores into two groups, *attack* and *victim* (Figure 3). This can be performed using the control group feature on Linux via the cset user-space management utility. We assign one core to the *attack* group, while putting all the remaining physical cores in the *victim* group. We then migrate all running processes to the *attack* group to minimize noise on the cores of the *victim* group. This will not always result in perfect idle situations, since migration can fail, e.g., for kernel threads. This means individual cores of the *victim* group may still contain more than one thread.

**Reducing Hardware Interference** Intel processors have mechanism deployed to ensure that thermal limits and power constraints are obeyed. In general, these mechanism play an active role in high-performance situations by reducing the P-state. To prevent interference at higher P-states, we disable them in our setup. Specifically, we disable the Thermal Control Circuit, Thermal Interrupt Control, PP0 and PP1 power limits as well as the package counterparts in the respective MSRs [28].

### 4.2 Undervolting x86 Processors

In the undervolting process the attacker searches for fault-prone voltage levels. Due to the shared voltage domain on x86-based platforms, we cannot target individual cores which makes containing faults within one core challenging (as opposed to, e.g., ARM-based platforms where fine-grained DVFS allows undervolting physical cores within their own voltage domain). Hence, our implementation makes use of a software-based approach

```
1  buffer[] input;
2  reference = algorithm(input);
3
4  // undervolting starts here
5  loop {
6    result = algorithm(input);
7    if (reference != result){
8      print_difference(reference, result);
9      exit;
10   }
11 }
```
Listing 1: Pseudo-code of our automated testing procedure.

```
1  _loop:
2    push %r10;
3    vpsllq %xmm3, %xmm4, %xmm6
4    vpsllq %xmm3, %xmm5, %xmm7
5    pop %r10;
6    jmp _loop;
```
Listing 2: Code of our most effective stressor.

which relies on two principles: *core isolation* and *selective probing*. Core isolation is established through our system setup as explained in the previous section. Selective probing means that only one *test* core is executing candidate programs while the system core increasingly undervolts and collects information about possible fault occurrences. Moreover, our setup establishes temperature differences between the cores. The idle cores have the lowest temperature. As the victim core is constantly executing code, it has the highest temperature. Additionally, we use *stressors* on the logical partner core to further increase the temperature. The temperature of the attack core lies between them. We want to keep this temperature as low as possible so the logical partner of the attack core is kept idle.

**Test Programs** We developed a set of test programs, which are based on the concept in Listing 1. The idea is to have conditional checks on deterministic results which stop execution when a deviation has been detected. First, we deterministically compute a reference result on Line 2. This step is conducted at normal operation voltage. Next, we execute the same computation but in a loop and using an undervolting setup. In each iteration we compare the reference output with the output of the previous iteration. Since the input is fixed and the target instructions perform deterministic operations on that input, any differences from the reference results indicates that a fault has corrupted the result.

**Stressors** In order to stress the undervolted components of the CPU, we looked for instruction sequences to execute on the logical partner of the target core.

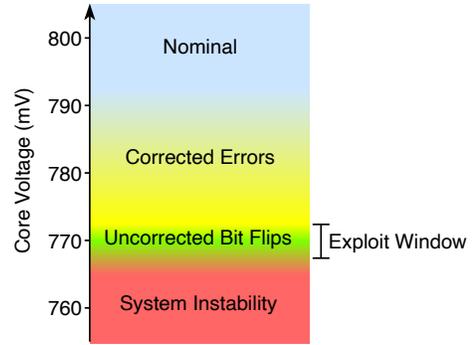

Figure 4: Processor behavior when exposed to reduced voltage. The voltage levels are only for illustrative purposes, since they vary according to processor model and P-state.

```
1    // logical vector operation
2    vpxor %xmm1, %xmm2, %xmm3
3    // data transfer to memory
4    vmovdqu %xmm3, (%rsp)
```
Listing 3: An instance of the vulnerable pattern VP1.

While the faults still happen frequently without stressors and even with hyperthreading disabled, we find that a good stressor improves the likelihood of faults. The best-performing stressor we found is in Listing 2. This stressor was deployed and running in all of our experiments.

**Fault Detection** In addition to the test programs, we relied on two more sources that indicated to us, when a fault occurred. First, the Machine Check Architecture (MCA), which delivers meta information about corrected and uncorrected faults in MSR. During our testing we monitored the respective MSR with existing tools like *mcelog*. For information about uncorrected errors, we were required to edit the MCE handler, either by dynamically instrumenting it or by compiling our own kernel.

Second, we monitor the operating system for processor exceptions like *Invalid Opcode* or *General Protection*. These exceptions might for instance be raised if the induced fault tampers with instruction decoding and therefore leads to the processor executing instructions that are not part of the correct code.

### 4.3 Bit flips in SIMD Memory Transfer

In Figure 4, we depict the observed behavior of the processor while it undergoes undervolting. As the voltage decreases, the processor starts to experience some errors that the MCA is able to correct (*Corrected Errors*). At a lower voltage, the system becomes unstable; the processor starts encountering hardware exceptions in interrupt handlers. However, between these two regions we encounter an *exploit window*, i.e., a voltage level where the processor experiences *uncorrected bit flips* that the MCA does not detect, but the system is still stable enough. In order to explore the exploit window, we implemented the concept in Listing 1 with common encryption algorithms like AES and Twofish. The programs continuously encrypt the same buffer and do not lead to faults under nominal voltage conditions. In our test setup, we executed the programs at different P-states while undervolting the core domain. We found some of the programs to be susceptible to faults when reaching specific voltage levels. This means that the comparison on Line 7 of Listing 1 revealed a difference in the computed results due to flipped bits in the output buffer. As depicted in Figure 4, the exploitable voltage level is located approximately 5 mV above the point, where the system starts to become unstable (e.g., due to exceptions in the kernel).

By manually analyzing the programs, we found the fault to affect two particular code patterns of SSE/AVX instructions:

**VP1** a parallel logic (e.g., xor) operation, followed by a move instruction from a vector register to memory, and

**VP2** a parallel add operation, followed by a move instruction from a vector register to memory.

An instance of the pattern VP1 is presented in Listing 3. On Line 2 of Listing 3, the exclusive OR (XOR) of registers xmm1 and xmm2 is computed and the result is stored in register xmm3. On Line 4 the value of this register is moved to memory, which in this case is indirectly addressed by a pointer in the register rsp.

## 5  Attacking SGX Enclaves

In the following we describe two different attack scenarios: first, our initial proof-of-concept attack that exploits bit flips inducing through undervolting in an enclave. Second, we present an attack against a real-world SGX crypto library developed by Intel.

### 5.1  From Bitflips to Attacks in SGX

We will now discuss how we leveraged the bit flips we discussed in Section 4.3 for the V0LTpwn attack. To illustrate the impact of bit flips on an SGX enclave, we start by considering some simple example code which first processes some input in memory and then branches execution based on the result. We provide a stripped down version of the relevant parts of the code in Listing 4, highlighting the most important parts in the form of inline assembly for clarity. Here, the enclave first performs a logical AND of the variables *a* and *b* through the instruction on Line 14. The enclave then copies the result

```
1  unsigned long a[2]={ULLONG_MAX, ULLONG_MAX};
2  unsigned long b[2]={ULLONG_MAX, ULLONG_MAX};
3
4  unsigned long r[2];
5
6  __asm__ __volatile__ (
7      "vmovdqu %1, %%xmm10;"
8      "vmovdqu %2, %%xmm11;"
9      "vpand %%xmm10, %%xmm11, %%xmm12;"
10     "vmovdqu %%xmm12, %0;"
11     :: "m" (*r) , "m" (*a), "m" (*b)
12     : "%xmm10","%xmm11","%xmm12", "memory");
13
14 if(r[0] == ULLONG_MAX && r[1] == ULLONG_MAX){
15     do_normal_operation();
16 } else {
17     do_recovery();
18 }
```

Listing 4: The enclave code used in our control-flow deviation PoC.

to the variable *r* on Lines 10 to 12. The variables a, b, and r represent a 128-bit vector encoding a particular program value, in this case ULLONG_MAX which causes every bit to be set to 1. Next, enclave execution checks the result against the ULLONG_MAX value on Line 14. In theory, this means that control flow should never reach Line 17 in this particular example. We would like to reiterate that this example code does not suffer from any software bugs and under normal circumstances the enclave execution *will always* take the branch on Line 15. However, using our fault injection attack we were able to force enclave execution into taking the else branch on Line 17 instead. We were able to perform this attack with up to 99% success rate: we provide detailed evaluation results about fault-inducing parameters and reliability of this particular exploit scenario in Section 6.3. Next, we are going to demonstrate how bit flips can be exploited in real-world SGX code.

### 5.2  Attacking Real-World SGX Code

Implementation of multiple cryptographic ciphers are especially prone to our fault injection attacks, including as OpenSSL and the crypto API of the Linux kernel. Hence, we demonstrate the feasibility of real-world V0LTpwn attacks by targeting an enclave running Intel's OpenSSL SGX library, which represents real-world crypto code that is specifically designed and intended to run inside an SGX enclave. We linked its latest Linux library version[2] against an enclave that validates a hash-based message authentication code (HMAC) using the cryptographic hash function SHA256.

We evaluated this attack on a Core i7-7700K and a Core i7-8700K processor. The microarchitecture of the former is Kaby Lake, that of the latter is Coffee Lake. We evaluated different message sizes and physical cores, while running the stressor from Listing 2 on the logical partner core. The results are summarized in Table 1, which reports the expected number of successes per 10 000 tries and the related standard deviation. For every core we conducted five independent test runs with two different message sizes (32B and 1KB). The adversary can reliably induce faults during hash computation on at least one physical core for each processor cores (namely core 1 for the 7700K and core 0 for the 8700K). An attacker utilizing these cores is able to induce faults in up to 34% of the HMAC validations on the 7700K and up to 99% on the 8700K. The other cores on the 7700K are unable to function at the same low voltage as core 1, while faults are rare at higher voltages. On the 8700K, cores 2 and 4 can function at the same low voltage as core 1 or even lower, but they only have a success rate of up to 30% and 16% respectively.

---
[2]Branch lin_2.5_1.1.1c of the repository at https://github.com/intel/intel-sgx-ssl.

| Processor | Core | Target core start temperature (°C) | Voltage (V) | Offset (mV) | 32B payload | 1KB payload |
|---|---|---|---|---|---|---|
| i7-7700K | 0 | 40 | 0.705 | -245 | 24.8 (σ=24.4) | 0.0 (σ=0.0) |
|  | 1 | 40 | 0.700 | -250 | **1795.6** (σ=1096.5) | **1983.8** (σ=364.2) |
|  | 2 | 40 | 0.710 | -240 | 821.2 (σ=321.0) | 745.2 (σ=148.8) |
|  | 3 | 40 | 0.710 | -240 | 283.6 (σ=119.9) | 235.2 (σ=51.6) |
| i7-8700K | 0 | 47 | 0.760 | -245 | **9621.6** (σ=146.7) | **9548.7** (σ=314.4) |
|  | 1 | 47 | 0.765 | -275 | 35.2 (σ=15.9) | 1320.2 (σ=243.3) |
|  | 2 | 47 | 0.755 | -285 | 2675.6 (σ=195.1) | 119.4 (σ=28.2) |
|  | 3 | 47 | 0.765 | -270 | 0.0 (σ=0.0) | 4.6 (σ=9.2) |
|  | 4 | 47 | 0.760 | -275 | 1496.8 (σ=148.1) | 1552.8 (σ=189.5) |
|  | 5 | 47 | 0.765 | -245 | 57.4 (σ=114.3) | 0.0 (σ=0.0) |

Table 1: Success rates of our attack to the OpenSSL HMAC implementation. For every core and payload size we report the expected successes per 10 000 tries and the related standard deviation (σ). We ran every test 5 times. In addition to absolute voltage levels, we present the offsets applied to MSR 0x150.[4]

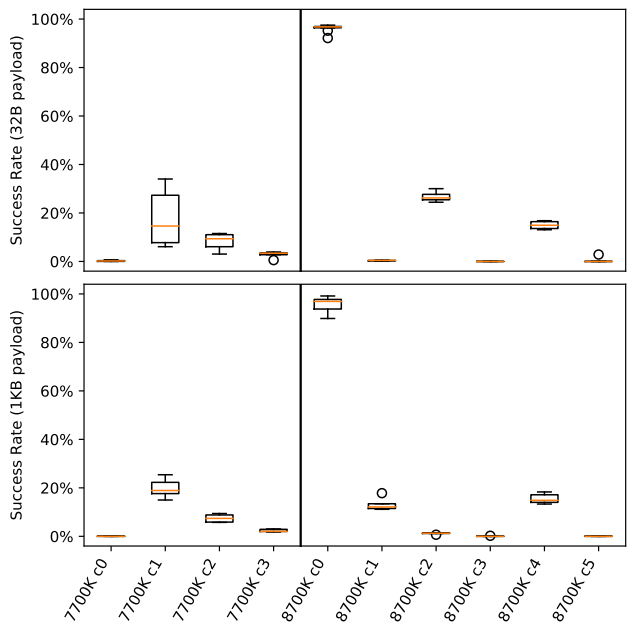

Figure 5: Success rate of our OpenSSL HMAC attack on various cores. The top graph refers to a payload size of 32B, the bottom one 1KB.

All in all, this shows that benign, real-world enclave code is susceptible to faults that can be provoked from software. This can be especially devastating from a security perspective during secret key generation. Since the computational security of public-key cryptography relies on the assumption that some mathematical problem is computationally hard, flipping a bit in one of the intermediate results could potentially weaken the security of the underlying cipher to enable real-world brute-forcing attacks.[5] Further, a number of recent works leverage TEEs to implement higher-level smart contract protocols [12, 13] or multi-party computation [18, 41]. Both of these use cases depend heavily on cryptography and we expect them to be highly affected by the V0LTpwn attack.

## 6 Evaluation and Results

In this section, we evaluate our V0LTpwn attack. In particular, we analyze at which voltage levels faults occur, how they manifest in memory (e.g., with respect

---
[4] In practice, we found that voltage offsets can vary slightly, as base voltage depends on a number of factors, such as active C-states, workload, as well as temperature.

[5] Further attack possibilities include denial of service when encrypting data, such that decryption becomes impossible due to a faulty key being used by the enclave.

to locality), and how reliably bit flips can be exploited within SGX.

## 6.1 Tested Platforms and Configurations

For the evaluation we used multiple Intel processors from different generations. In detail, we used the i7-7700 and i7-7700K with the Kaby Lake microarchitecture and the i7-8700K from the Coffee Lake generation[6].

We conducted preliminary testing on these platforms which we found to be prone to non-recoverable, software-induced processor faults due to undervolting. Our platforms are running the official Intel SGX SDK, PSW and drivers in version 2.5 released in May 2019 for Ubuntu 18.04 (minimal installation).

We created an example SGX enclave which we build in Hardware-PreRelease mode.

## 6.2 Fault-Inducing Voltage Level

To demonstrate that bit flips can be reproduced at arbitrary P-states, we evaluated the set {0x8, 0x10, 0x1B, 0x20, 0x24, 0x2A} on our test processors. We used the same setup as described in Section 4 and executed a program containing the vulnerable code pattern (Listing 3) on every core. For every run, we measured the earliest fault-prone voltage level. In Table 2, we present the results for the i7-7700K processors. In general, we observe that every P-state has custom fault-prone voltage levels. Depending on the P-state, the voltage offset, which has to be applied to MSR 0x150, ranges between 250 mV and 300 mV. For every P-state, we measure differences of 5 to 10 mV between the cores.

Repeating the same procedure on the other processors yields the same observations. However, every processor model has individual fault-prone voltage levels. We assume the cause lies in variations in the manufacturing process. Regarding the V0LTpwn attack, the result implicate that an attacker has to adapt the attack parameters for every target processor.

## 6.3 Evaluation of the Control-flow Deviation PoC

We evaluated our proof-of-concept control flow deviation exploit (described in Section 5.1) on all cores of our i7-7700K processor, spanning the whole range of available P-states. We created an SGX enclave which runs the code in Listing 4 10 000 times. We then tried running the enclave in various undervolted environments for 100 000 times. Figure 6 shows the success rate of the attack, i.e., the percentage of runs in which the different branch was executed in Listing 4. We tested two different stressors: the stressor from Listing 2 and an AVX implementation of the Twofish cipher [27]. The best-performing stressor is the code from Listing 2; while using this stressor, cores 1, 2, and 3 achieved success rates of 99%, 96% and 99% respectively at 700 mV and P-state 0x1B. Using the Twofish code as a stressor, we could only achieve up to 8% success rate on core 1 at P-state 0x1B, 6% at P-state 0x20, and 2.5% on P-state 0x24. Cores 2 and 3 reached a success rate of 2.5% Core 0 did not show a significant number of faults.

---

[6] Intel uses Stepping codes to differentiate between different revisions of a microarchitecture. Our Kaby Lake processor has Stepping 9 and our Coffee Lake has Stepping 10.

| P-state | T (°C) | Core 0 | Core 1 | Core 2 | Core 3 |
|---|---|---|---|---|---|
| 0x08 | 32 | 0.540 | 0.545 | 0.535 | 0.545 |
| 0x10 | 33 | 0.585 | 0.585 | 0.580 | 0.585 |
| 0x1B | 37 | 0.700 | 0.710 | 0.705 | 0.705 |
| 0x20 | 41 | 0.765 | 0.775 | 0.770 | 0.775 |
| 0x24 | 42 | 0.825 | 0.835 | 0.835 | 0.835 |
| 0x2A | 50 | 0.930 | 0.935 | 0.930 | 0.935 |

Table 2: Fault-prone voltage levels (V) for different P-states and cores of i7-7700K processor.

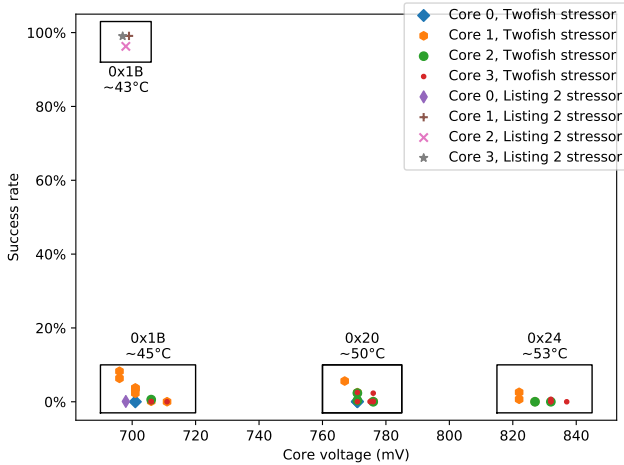

Figure 6: Reliability results of our proof-of-concept on the i7-7700K processor: success rate of the PoC exploit in Listing 4.

We could only obtain faults in P-states between 0x1B (2700 MHz) and 0x24 (3600 MHz). Lower P-states yield practically no successful attacks but only lead to recoverable errors. This is reasonable, since a lower P-state effectively means that the processor is running at a lower frequency (e.g., 800 MHz for P-state 0x8 and 1600 MHz for P-state 0x10), and hence, requires overall less power to execute instructions. Therefore, lowering the voltage supply is not an effective measure to produce faults on the lower frequency domain — at least not within the limits available from software. Pushing the system towards the high frequency limits did not produce better exploit reliability after a certain point. While perhaps counter-intuitive at first, this can be explained by two facts: first, higher frequency domains naturally require higher voltage levels. This means that the base voltage that is supplied to the cores in that state will be higher. However, the voltage offset the attacker is able to set to reduce the voltage supply from software is limited to a fixed range, and hence, affecting core voltage from software in this way is less effective in the higher frequency domain. Second, it has been known for a long time that hardware becomes generally less stable as clock frequency increases [49]. This means, any physical effect interfering with normal processor execution has more severe consequences for the overall system at higher clock frequencies. For instance, in our tests we observed that the system will more easily crash in a hard crash than produce machine-check exceptions.

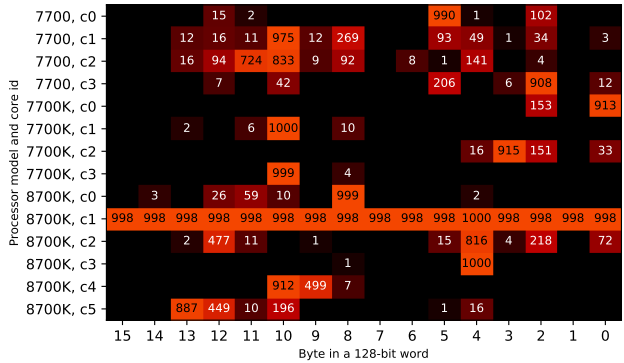

Figure 7: Heat map of the location of bit flips inside a 128-bit word, for 1000 faults on each core of each processors.

| Processor | Core | 1 BF | 2 BF | 3+ BF |
|---|---|---|---|---|
| i7-7700 | 0 | 905 | 83 | 12 |
|  | 1 | 709 | 199 | 92 |
|  | 2 | 405 | 444 | 151 |
|  | 3 | 855 | 122 | 23 |
| i7-7700K | 0 | 934 | 66 | 0 |
|  | 1 | 988 | 7 | 5 |
|  | 2 | 912 | 67 | 21 |
|  | 3 | 997 | 3 | 0 |
| i7-8700K | 0 | 942 | 32 | 26 |
|  | 1 | 2 | 0 | 998 |
|  | 2 | 589 | 275 | 136 |
|  | 3 | 999 | 1 | 0 |
|  | 4 | 586 | 410 | 4 |
|  | 5 | 614 | 239 | 147 |

Table 3: Breakdown of 1000 faults on various cores and processors: for every core, the table shows how many faults led to one bit flip, two bit flips, and three or more bit flips

### 6.4 Fault Manifestation

Being able to induce faults in a reproducible way from software allowed us to study the behavior and details behind the generated faults. We analyzed the faults with regards to their position our three processors: i7-7700, i7-7700K, and i7-8700K. We made several interesting observations: first, all faults we observed manifested as bit flips in the result of computation or memory transfers. Second, bit flips affected different byte positions within the respective 128-bit word used by the faulting instructions (Figure 7). Since the minimal, vulnerable instruction patterns VP1 and VP2 utilize vector operations, we focused on 128-bit words used by AVX instructions in our subsequent analysis. Our tests show that faults are significantly more likely for certain byte positions, while other locations were never affected. The affected bytes are different for each physical core we tested: for instance, on core 3 of the 8700K faults were heavily localized within byte 4, while the remaining cores were affected by bit flips throughout several different byte positions. In contrast to this, core 1 was affected by bit flips within *all* byte position. Interestingly, the number of bit flips produced per fault also varied between cores (Table 3). On the 7700K, physical cores were likely to yield only a single bit flip, while on the 7700 we observed a larger number of multi-bit errors. On the 8700K, we observed both single-bit and multi-bit faults.

Perhaps most crucially, the affected byte locations remained stable for a given physical core: the bit flip positions were reproducible on each core at different times and also consistent across different P-states.

## 7 Discussion

Being able to compromise the integrity of computations is a powerful tool in the hands of software adversaries. So far, we were able to confirm successful fault-injection attacks from software against certain vulnerable code patterns, which have to be part of the victim code (Listing 3). These susceptible pieces of code we identified are naturally used in many implementations, e.g., to optimize the performance using SIMD instructions. We also conducted another series of tests using non-temporal instructions, such as `movnti` and `movntq` followed by an `sfence` instruction as replacement. These non-temporal instructions bypass the caches and access memory directly. Our results showed that we still were able to achieve reproducible bit flips and the patterns did not change due to non-temporal move instructions. We conclude that bit flips in the result must have been intro-duced by the physical core as opposed to one of the caching structures, e.g., execution units, the register file, read or write buffers, or possibly one of the buses.

In our analysis we identified the respective, susceptible vector operations in many real-world implementations of cryptographic algorithms. As we demonstrate, we were able to exploit these fault-susceptible instruction patterns to achieve memory corruption *in the absence of software vulnerabilities* by undervolting the processor.

In this paper we demonstrated attacks against SGX enclaves, however, other attack scenarios might be viable within our threat model. For instance, an adversary might try to break Mandatory Access Control on SELinux [38]. Further, during our testing we noticed that the voltage setting through MSR 0x150 remains in place after rebooting the system (i.e., through *warm reset*). This opens up the possibility of targeting bootloader code, which typically is the root of trust on modern platforms.

Another interesting aspect is that we occasionally observed the *Invalid Opcode* processor exception while undervolting our testing code. This exception is usually raised if the processor encounters a malformed instruction. However, since our testing code only contained valid, well-formed instructions, this exception must have been introduced by our undervolting. The MCA logs confirmed this observations by reporting *instruction decode corrected errors*, leading us to conclude that it is possible to tamper with instruction decoding through undervolting in principle. However, we leave an in-depth investigation of this to future work.

Currently, our attack focuses on Intel processors (which deploy SGX) and we did not test or evaluate our attack on AMD systems. While confidentiality of Intel processors has been attacked in many prior publications, V0LTpwn is—to the best of our knowledge—the first successful attack on processor integrity for x86.

## 8 Related Work

As explained earlier, V0LTpwn constitutes the first remote-fault injection attack for the x86 platform. However, related attacks have been presented for other platforms and a number hardware-oriented side-channel attacks were published recently for x86 which do not involve fault injection. In this Section we first elaborate how V0LTpwn compares to existing work and attacks that were presented previously. Second, we present a quick overview of the related tools and methods for conducting fault-injection attacks from software.

### 8.1 Hardware-Oriented Exploits

For a direct comparison, we only focus on hardware attacks that are within the scope of our threat model, i.e., attacks that do *not* require physical presence but can be launched *remotely* from software.

#### 8.1.1 Software-Controlled Fault Injection

The CLKScrew [53] attack first demonstrated that sophisticated power-management APIs on some ARM-based devices allow an adversary to induce faults in the processor entirely remotely. These findings were recently reproduced independently by VoltJockey [43]. In both cases, the authors were able to break the TrustZone isolation boundary on a Nexus 6 smartphone. Unfortunately, the techniques used to conduct undervolting attacks on ARM are not transferable to x86-based platforms for several reasons: first, both Tang et al. and Pengfei et al. found core voltage and frequency to be exposed directly to software, with practically no limitations or restrictions imposed by the ARM architecture. This means, the attacker is able to freely chose practically arbitrary combinations of frequency and voltage pairs, allowing them to construct and apply utterly unsafe settings entirely from software to conduct their attack. By contrast, the x86 platform offers only a fixed, pre-defined list of selected p-states that are extensively tested for their safety margins and common operating conditions by the manufacturer prior to release. Hence, the attacker is constrained to use one of these hand-picked frequency voltage pair definitions to conduct a V0LTpwn.

Second, Intel deploys the Machine-Check Architecture to explicitly check for and recover from hardware faults at run time. Since Machine-Check Exceptions originating from any core are broadcast to all cores, certain hard glitches can effectively be converted into soft errors on-the-fly on x86 and our evaluation shows that the attacker has to push the victim core beyond a certain threshold to ensure successful faults and exploitation. Further, individual hardware components such as the caches and the core have to be undervolted in lock-step for any changes to take effect on x86. This means that faults generated from any other of these other components contribute to the early warning mechanism employed by the Machine-Check Architecture. No such safety net exists on ARM, significantly facilitating reliability of faults and reproducible exploit scenarios.

Third, the core pinning technique introduced by Tang et al. [53] ensures that faults are contained within a chosen physical core, making it straightforward to launch attacks against a target core from one of the running system cores as an attacker. This technique works since each core can effectively operate in its own p-state on ARM. On x86 all physical cores operate within the same p-state, which means that the same voltage settings apply to the attacker as well as the victim core, and hence, faults cannot simply be contained to any given core. This is why we introduce several novel techniques to ensure an overall stable system while being able to force the victim core into a fault-provoking power domain on x86.

Lastly, since power-management is one of the key driving factors on mobile devices the related low-level APIs and involved hardware mechanisms are extensively documented and tooling is readily available, or even built into the existing platform software [43, 53]. On x86 practically no official documentation regarding low-level power management of the platform exists, making it hard to develop custom tools and even conducting simple tests usually involves costly reverse engineering of micro-architectural features, which can also differ between the many processor generations.

### 8.1.2 Rowhammer

Rowhammer attacks [32] are similar in nature to CLKScrew [53] and V0LTpwn in so far as they generate hardware faults from software that are also exploitable [6, 26, 31, 42, 46, 48, 54, 57, 60]. However, the main difference from our work is that Rowhammer affects DRAM, which is widely used for implementing the memory modules on off-the-shelve computing hardware. This means Rowhammer attacks cannot affect memory inside the processor, such as cached memory and register values. In contrast to this we show that V0LTpwn directly impacts in-processor values and can also divert control flow. Additionally, while several countermeasures [5, 8, 54, 58] have been proposed to mitigate Rowhammer from software, no defenses currently exist to counter processor-based fault injection attacks.

### 8.1.3 Speculative Execution

Recently, several works independently demonstrated that speculative execution (a processor feature to speed up execution by increasing instruction-level parallelism) could be exploited from software on certain platforms to extract information through a side channel [33, 36, 55, 59]. Unlike attacks based on speculative execution (some of which can be addressed fully through software-only defenses [19, 25]) remote-fault injection attacks are not limited to information disclosure, but directly affect system integrity, allowing an adversary to *manipulate* data as well as execution.

## 8.2 Analyzing x86 Internals

Earlier work by Pandit et al. [39] analyzed voltage offsets with regards to safe operation limits, with a focus towards increased processor performance. In that context, they analyzed error handling of the *Machine Check Architecture* on AMD processors and found that during undervolting they were able to operate it beyond safe operation points. They also observed corrected machine check errors when reaching a threshold voltage offset and showed an increased error rate at higher CPU utilization.

Another study by Papadimitriou et al. [40] investigated voltage offsets on mobile and desktop processors from Intel. They used standard benchmarks to stress cores while applying voltage offsets with Intel's XTU application and found that voltage can be decreased up to 15% while keeping the system in an overall stable condition. They observed differences in safe voltage offsets for the analyzed processor models and calculated that safe undervolting can lead to an increased energy-efficiency of up to 20% and temperature reductions of up to 25%.

More recently, Koppe et al. [34] presented a framework to analyze as well as synthesize x86 microcode on certain (older) platforms. Christopher Domas presented initial results on reverse engineering the x86 hardware platform and published several tools [15, 16] to automatically uncover certain aspects and features (including undocumented MSRs). Domas also discovered hardware backdoors through hidden modes on certain VIA x86 processors using those tools.

Researchers from Positive Technologies achieved remote code execution on Intel's Management Engine (Intel ME) in 2018 [20]. Intel ME runs on a separate physical chip from the main host CPU that remains powered even if the main CPU is in deep sleep. Intel ME has full platform access, drives all security-related tasks on modern Intel platforms (including SGX, TXT, AMT) and was recently found to even include a full-blown logic analyzer called Intel VISA [21]. This already offered great insight into how Intel patched hardware vulnerabilities in microcode as well as uncovering previously unknown platform internals. In the future Intel VISA could have the potential to unlock even more proprietary information about the x86 platform as well as aid in reversing and debugging micro-architectural features such as MCA.

## 9 Conclusions

In this paper we introduced V0LTpwn, a novel software-controlled fault-injection attack that leverages frequency and voltage control interfaces to compromise the integrity of x86 processors. We find and discuss multiple code patterns that are prone to bit flips and are commonly used in crypto code. We show that V0LTpwn can generate faults in real-world OpenSSL code running in an SGX enclave with a success rate of up to 99%. We analyze the success rate of V0LTpwn over a variety of parameters.

| **Domain** [42:40] | | **Command** [39:32] | |
|---|---|---|---|
| 0x0 | Cores | 0x10 | Read Voltage Change |
| 0x1 | Core GPU | 0x11 | Write Voltage |
| 0x2 | LLC/Ring | | |
| 0x3 | System Agent | | |

Table 4: Relevant domain and command encodings for using MSR OC Mailbox (`0x150`) from software.

## A  OC Mailbox Interface

In Table 4 we list the possible domain and command encodings that are known to us. Not all x86 platforms are designed for overclocking, so the commands may not be available on all systems. However, we found the voltage read/write commands `0x10`/`0x11` to be present in all newer mobile and desktop platforms. The write command is used to modify the voltage of the domain unit and is present in the two modes offset (`0x`) and static (`0x1`), which can be selected by bit [20] of the payload. The offset mode applies the offset value located in the bits [31:21] to the voltage of the domain. The offset is encoded as an 11 bit signed value, allowing a theoretical offset range from -1024 mV to 1023 mV. For the domain Core (`0x0`), the offset is applied to the base voltage of every p-state. As an example, writing the value `0x80000011f3800000` to the OC Mailbox MSR, will apply an offset of -100 (`0xf38`) mV to every p-state.

In static mode, the domain voltage can be set to a fixed value that is encoded in the bits [19:8] of the payload. This 11 bit unsigned value is divided by 1024 by the hardware, allowing to set a static voltage from 0 to 2V. In the rest of the paper, only the offset mode is used to control the voltage. When we use the term *undervolting* we mean applying a negative offset via this command through the MSR OC Mailbox (`0x150`). We would like to emphasize again that any details related to MSRs can in principle depend on the micro-architectural generation and model version of the processor.

16